\documentstyle[12pt]{article}
\def\slash#1{\setbox0=\hbox{$#1$}#1\hskip-\wd0\hbox to\wd0{\hss\sl/\/\hss}}

\begin{document}

\begin{flushright}
MZ-TH/92-49 \\
October 1992
\end{flushright}
\bigskip

\begin{center}
{\bf{\LARGE Leptonic Flavor-changing {\boldmath $Z^0$} Decays}} \\[0.45cm]
{\bf{\LARGE in {\boldmath $SU(2) \otimes U(1)$} Theories }}\\[0.45cm]
{\bf{\LARGE with Right-handed Neutrinos}} \\[1.5cm]
\bigskip\bigskip\bigskip
{\large J.~G.~K\"orner \footnote[1]{work supported in part by the BMFT,
{\em FRG}, under contract 06MZ730.},
A.~Pilaftsis \footnote[2]{supported by a grant
from the Postdoctoral Graduate College of Mainz.},
K.~Schilcher \footnotemark[1] } \\
 Institut f\"ur Physik \\
 Johannes-Gutenberg-Universit\"at \\
 Staudinger Weg 7, Postfach 3980 \\
 W-6500 Mainz, {\em FRG}
\end{center}

\bigskip
\centerline{to appear in {\em Phys.~Lett.~B}}
\bigskip\bigskip
\centerline {\bf ABSTRACT}

We analyze possible lepton-flavor-violating decays of the $Z^0$
particle in a minimal extension
of the Standard Model, in which one right-handed neutral field for each
family has been introduced. Such rare leptonic decays are induced by
Majorana neutrinos at the first electroweak loop level and are generally
not suppressed by the ordinary "see-saw" mechanism. In particular,
we find that experimental bounds on branching ratios of the order of
$10^{-5}-10^{-6}$ attainable at $LEP$ may impose constraints on
lepton-flavor-mixing parameters and the masses of the heavy Majorana
neutrinos.\\

\newpage

\newpage

Many theoretical scenarios have recently been proposed for the production
and the lepton-number-violating decay of heavy Majorana neutrinos at
accessible accelerator energies~[1,2,5]. A low-energy realization of these
theories could be the Standard Model ($SM$) with one right-handed neutrino
field per family~[4]. Such a three-generation "see-saw" model~[3] can
theoretically provide small Majorana masses for the ordinary
neutrinos, relatively light masses for the heavy ones (e.g. of the order
of 100~GeV), as well as large light-heavy neutrino mixings (i.e.
$\xi_{\nu N} \simeq 0.1$)~[2]. This possibility was systematically investigated
in~[5] and the notation given there will be used throughout this note.
In this minimal class of models some interesting phenomenological
applications to possible lepton-flavor-violating $H^0$ decays have recently
been  presented~[6].
In this note, however, we will calculate branching ratios of
leptonic flavor-changing decays of the $Z^0$ particle, i.e. $Z^0 \to e\tau$ or
$\mu\tau$, in this minimal class of models. For realistic samples of
$10^6-10^7$~$Z^0$ decays at $LEP$,
one would be sensitive to branching ratios of the order of $10^{-5}-10^{-6}$.
We find that branching ratios of this order of magnitude could be easily
understood within such minimal models.
Of course, alternative theoretical models exhibiting
lepton-flavor-nonconservation in $Z^0$ decays, have been
proposed over the years~[7,8]. We briefly comment on the results of some
of these calculations.\\

As has been extensively discussed in~[5], in the Feynman-'t Hooft gauge
the relevant couplings of Majorana neutrinos $n_i$ (with $i=1,2\dots,
2n_G$) to the $W^\pm$, $Z^0$ and Goldstone bosons $\chi^\pm$
are written down in terms of mass eigenstates as follows:
\begin{eqnarray}
{\cal L}_{int}^W& = & -\frac{g_W}{2\sqrt{2}} W^{-\mu}\
{\bar{l}}_i \ B_{l_ij} {\gamma}_{\mu} (1-{\gamma}_5) \ n_j \quad + \quad h.c.
\\[0.3cm]
{\cal L}_{int}^Z& = & -\frac{g_W}{4\cos\theta_W}  Z^{0\mu}\
\bar{n}_i \gamma_\mu [ i\mbox{Im}(C_{ij})\ -\ \gamma_5\mbox{Re}(C_{ij}) ]
n_j\\[0.3cm]
{\cal L}_{int}^{\chi}\  &=&  -\frac{g_W}{2\sqrt{2}M_W} {\chi}^-\
{\bar{l}}_i \ [m_{l_i}B_{l_ij} (1-{\gamma}_5)\ -\ B_{l_ij}(1+{\gamma}_5)
m_{n_j}] \ n_j\quad + \quad h.c.
\end{eqnarray} 
where $B_{l_ij}$ is a rectangular $n_G\times 2n_G$ matrix given by
\begin{equation}
B_{l_ij} = \sum\limits_{k=1}^{n_G} V^l_{l_ik} U^{\nu\ast}_{kj}
\end{equation} 
and $C_{ij}$ is in general a nondiagonal $2n_G\times 2n_G$ projection
matrix defined by
\begin{equation}
C_{ij} = \sum\limits_{k=1}^{n_G} U^{\nu}_{ki} U^{\nu\ast}_{kj}
\end{equation} 
In eqs~(4) and~(5), $V^l$ and $U^\nu$ are the corresponding
Cabbibo-Kobayashi-Maskawa~($CKM$) matrix for the leptonic sector and the
unitary matrix which diagonalizes the symmetric "see-saw" neutrino mass
matrix, respectively.\\

In this low-energy model, Majorana neutrinos can give rise to
flavor-changing neutral current effects in the leptonic sector at the first
electroweak loop level. The Feynman diagrams responsible for the decay
process $Z^0\to l_1\bar{l}_2\ \ (l_1\neq l_2)$ are shown in figs (1a)--(1k).
Before we proceed to list their individual contributions to the above
leptonic rare decays, we first introduce some useful abbreviations:
\begin{eqnarray}
{\mbox{C}}_{UV} & = & \frac{1}{\varepsilon} - {\gamma}_E + \ln 4\pi -
\ln \frac{M^2_W}{\mu^2} \\
{\Delta}_{ij} & = & \frac{ig_W{\alpha}_W}{8\pi\cos \theta_W}
\ B_{l_1i}B_{l_2j}^{\ast}
\ \ {\bar{\mbox{u}}}_{l_1} \gamma_\mu (1-\gamma_5) {\mbox{v}}_{l_2}
\ \varepsilon^\mu_Z\\
{\lambda}_i &=& \frac{m_{n_i}^2}{M^2_W} \ , \qquad
\lambda_Z =\frac{M^2_Z}{M^2_W}
\end{eqnarray} 
We also work in the massless limit for the external leptons.
Then, the results for the different contributing graphs to the decay
$Z^0\to l_1\bar{l}_2$ are obtained in the on-shell renormalization scheme~[9]
by the following expressions:
\begin{eqnarray}
i\Gamma_a & = & \ \frac{1}{2} \Delta_{ij} \left\{ C_{ij}
\left[\  \int\limits_0^1 dx dy \ y\ln B_2(\lambda_i,\lambda_j)\ - \
\lambda_Z \int\limits_0^1 \frac{dx dy\ y}{B_2(\lambda_i,\lambda_j)}
[1-y+y^2x(1-x)] \right] \right. \nonumber\\
&& \left. + \ C^\ast_{ij}\sqrt{\lambda_i\lambda_j} \int\limits_0^1
\frac{dxdy\ y}{B_2(\lambda_i,\lambda_j)} \right\}\\[0.4cm]
i\Gamma_b & = & -\ \frac{1}{4} \Delta_{ij} \left\{
C_{ij}\lambda_i\lambda_j \int\limits^1_0 \frac{dxdy\ y}
{B_2(\lambda_i,\lambda_j)}\  +\  C^\ast_{ij} \sqrt{\lambda_i\lambda_j}
\Bigg[\  \frac{1}{2}\mbox{C}_{UV}\ -\ \frac{1}{2} \right.\nonumber\\
&&\left. +\ \lambda_Z\int\limits_0^1 \frac{dxdy\ y^3x(1-x)}
{B_2(\lambda_i,\lambda_j)}\ - \ \int\limits_0^1 dxdy\ y\ln
B_2(\lambda_i,\lambda_j) \Bigg] \right\} \\[0.4cm]
i\Gamma_c & = & -\ \Delta_{ii} \left[ \lambda_Z \int\limits_0^1
\frac{dxdy}{B_1(\lambda_i)} y^2[1-yx(1-x)] \ +\  3\cos^2\theta_W
\int\limits_0^1 dxdy\ y\ln B_1(\lambda_i) \right]\\[0.4cm]
i\Gamma_d &=& \frac{1}{8}\Delta_{ii}(1-2\sin^2\theta_W)\lambda_i
\left[ \ \mbox{C}_{UV}\ -\ 2\int\limits_0^1 dxdy\ y \ln
B_1(\lambda_i) \right]
\end{eqnarray} 

\newpage

\begin{eqnarray}
i\Gamma_e+i\Gamma_f & = & -\ \Delta_{ii}\ \frac{\sin^2\theta_W}
{\cos\theta_W} \ \lambda_i \int\limits_0^1 \frac{dxdy\ y}
{B_1(\lambda_i)} \\[0.4cm]
i\Gamma_g+i\Gamma_j & = & \frac{1}{4} \Delta_{ii} (1-2\sin^2\theta_W)
\left[ \ \frac{\lambda_i}{1-\lambda_i}\ +\ \frac{\lambda_i^2 \ln
\lambda_i}{(1-\lambda_i)^2}\ \right]\\[0.4cm]
i\Gamma_h+i\Gamma_k & = & -\ \frac{1}{8}\Delta_{ii}
(1-2\sin^2\theta_W)\ \lambda_i \left[ \
\mbox{C}_{UV}\ +\ \frac{3}{2}\ -\ \frac{1}{1-\lambda_i}\ -\
\frac{\lambda_i^2\ln\lambda_i}{(1-\lambda_i)^2} \right]
\end{eqnarray} 
Above, the summation convention is assumed for repeated indices, which run
over all the Majorana neutrino mass-eigenstates $n_i$. Also, the functions
$B_1(\lambda_i)$ and $B_2(\lambda_i,\lambda_j)$ appearing in eqs~(9)--(13)
are defined as
\begin{eqnarray}
B_1(\lambda_i) &\  = \ & (1-y)\lambda_i + y[1-\lambda_Z\; yx(1-x)] \\
B_2(\lambda_i,\lambda_j) &\ = \ & 1-y+y[x\lambda_i+(1-x)\lambda_j-
\lambda_Z\; yx(1-x)]
\end{eqnarray} 
Notice that the $UV$ divergences are mass-dependent expressions and vanish
when all the diagrams are added together. To make this explicit,
we list the following useful identities~[5,6]:
\begin{eqnarray}
\sum\limits_{i=1}^{2n_G} B_{l_1i}B_{l_2i}^{\ast} & = & \ {\delta}_{l_1l_2}\\
\sum\limits_{i=1}^{2n_G} B_{li}C_{ij} & = & \ B_{lj} \\
\sum\limits_{i=1}^{2n_G} m_{n_i} B_{li}C^{\ast}_{ij} & = & \ 0 \\
\sum\limits_{k=1}^{n_G} B_{l_ki}^{\ast}B_{l_kj} & = & \ C_{ij} \\
\sum\limits_{i=1}^{2n_G} m_{n_i}B_{l_1i}B_{l_2i} & = & \ 0
\end{eqnarray} 
Analytically, eq.~(18) represents the generalized version of the $GIM$
mechanism which results in the cancellation of the mass-indepedent
$UV$ divergences in the graphs~(1a), (1c), (1g) and (1j).
The diagram~(1d) cancels the $UV$ pole of (1h) and (1k),
while the $UV$~constant in eq.~(10) vanishes due to eq.~(20). As a result
of these cancellations,
the transition amplitude for $Z^0\to l_1\bar{l}_2$ will be finite.
Another consequence of the identities~(18)--(22) is that they enormously
reduce the number of independent mixing parameters $B_{lj}$ and $C_{ij}$.
Thus, in our numerical analysis we make use of the fact that
\begin{eqnarray}
B_{l_1i}B_{l_2i}^\ast \ F(\lambda_i) & = &
B_{l_1N_i}B_{l_2N_i}^\ast\ \left[ F(\lambda_{N_i})-F(0) \right] \\
B_{l_1i}C_{ij}B_{l_2j}^\ast\ G(\lambda_i,\lambda_j) & = &
B_{l_1N_i}B_{l_2N_j}^\ast \Big\{ C_{N_iN_j}  \Big[
G(\lambda_{N_i},\lambda_{N_j})-G(\lambda_{N_i},0)-G(0,\lambda_{N_j})
\nonumber\\
&& +G(0,0) \Big] \ +\ \delta_{N_iN_j} \Big[ G(\lambda_{N_i},0)+
G(0,\lambda_{N_i})-G(0,0) \Big] \Big\}
\end{eqnarray} 

\indent

For the sake of illustration, we shall restrict our
discussion to a two generation model in the following.
Employing eqs~(18)--(22), we find the
following useful relationships among the different $CKM$-mixing combinations:
\begin{equation}
B_{l_1N_2}B_{l_2N_2}^\ast\ = \ \frac{m_{N_1}}{m_{N_2}}
B_{l_1N_1}B_{l_2N_1}^\ast\ , \quad\
B_{l_1N_2}C_{N_2N_2}B_{l_2N_2}^\ast\ =\ \frac{m^2_{N_1}}{m^2_{N_2}}
B_{l_1N_1}C_{N_1N_1}B_{l_2N_1}^\ast
\end{equation}
\begin{eqnarray}
B_{l_1N_1}C_{N_1N_2}^\ast B_{l_2N_2}^\ast\ &=&\
B_{l_1N_2}C_{N_2N_1}^\ast B_{l_2N_1}^\ast\ =\ -\frac{m_{N_1}}{m_{N_2}}
B_{l_1N_1}C_{N_1N_1}B_{l_2N_1}^\ast \\[0.4cm]
B_{l_1N_1}C_{N_1N_2} B_{l_2N_2}^\ast\ &=&\
B_{l_1N_2}C_{N_2N_1} B_{l_2N_1}^\ast\ =\  \frac{m_{N_1}}{m_{N_2}}
B_{l_1N_1}C_{N_1N_1}B_{l_2N_1}^\ast
\end{eqnarray} 
In order to pin down numerical predictions, we can estimate
$C_{N_1N_1}$ by using Schwartz's inequality in eq.~(21), namely
\begin{equation}
C_{N_1N_1} \ \ \geq \ \ 2|B_{l_1N_1}B_{l_2N_1}^{\ast}|
\end{equation} 
Then, in this illustrative two generation model we are left with only one
free mixing parameter; the $CKM$-mixing
combination $\zeta = B_{l_1N_1}B_{l_2N_1}^{\ast}$. In the $e-\mu$ system,
this quantity is strongly constrained to be $\zeta \leq 3\ 10^{-3}$ from
the non-observation of the decay mode $\mu \to e\gamma$~[10].
Constraints referring to $e-\tau$ or $\mu -\tau$ system are,
however, much weaker  and allow for mixing values
of $\zeta \leq 10^{-1}$~[11].\\

In the heavy neutrino limit (e.g. $m_{N_1} \simeq m_{N_2} \simeq m_N \gg
M_Z$) the amplitude ${\cal A}$ for the lepton-flavor-violating decay of the
$Z^0$ behaves like
\begin{equation}
{\cal A}(Z^0\to l_1\bar{l}_2)\ \simeq\ -\
\frac{ig_W\alpha_W}{16\pi\cos \theta_W}
\ \zeta^2 \ \frac{m^2_N}{M^2_W}\
\bar{\mbox{u}}_{l_2} \gamma_\mu(1-\gamma_5)\mbox{v}_{l_1}\ \varepsilon^\mu_Z
\end{equation} 
{}From this we can get a first estimate for the branching ratio of the rare
decays,
\begin{equation}
Br(Z^0 \to l_1\bar{l}_2+\bar{l}_1l_2) \ \simeq\  \frac{\alpha^3_W}
{192\pi^2\cos^2\theta_W} \ \frac{M_Z}{\Gamma_Z}\ \zeta^4\
\frac{m_N^4}{M^4_W}
\end{equation} 
where ${\Gamma_Z}$ denotes the total decay width of the $Z^0$~boson, given
by the central value ${\Gamma_Z}=2.534$~GeV~[12]. For a given fixed value
of $\zeta$, eq.~(30) shows a dramatic fourth power depedence of
the heavy neutrino mass $m_N$. In table~1 we present the numerical results
for an exact numerical computation of the branching ratios
$Br(Z^0 \to l_1l_2)$, using a value
of $\zeta=0.05$. We can readily see that experimental bounds on
$Br(Z^0\to e\tau, \mu\tau)$ of the order of $10^{-5}-10^{-6}$~[12]
lead to combined constraints on the lepton-flavor-mixing parameter
$\zeta$ and the heavy neutrino mass~$m_N$. In fig.~(2) we show the
results of such an analysis for two nearly degenerate neutrinos. However,
we have also checked that the numerical results do not show any essential
change when one considers large mass differences between the two
heavy Majorana neutrinos. In that case, $m_N$ in fig.~(2) indicates the
mass of the lightest heavy Majorana neutrino.\\

At this point an additional remark on the heavy neutrino masses
is in order. For a fixed value of the mixing parameter $\zeta$,
the heavy neutrino mass cannot be arbitrarily large, since $m_N$ is contrained
by the requirement of pertubative unitarity. We can safely estimate
this unitarity bound by imposing the inequality condition
$\Gamma_N/ m_N \leq 1/2$, where $\Gamma_N$ is the width of the heavy neutrino
$N$. As calculated in~[5], we obtain the inequality
\begin{equation}
m_N \ \ \leq \ \ 2M_W \sqrt{\frac{1}{\alpha_W \zeta}}
\end{equation} 
It should be also emphasized that, unlike the leptonic
flavor-changing $Z^0$ decays, the rare decays $\mu \to e\gamma$ or
$\tau \to \mu\gamma$ show a constant behavior with
increasing heavy neutrino mass in most theories.
This can be attributed to the fact that
the Feynman graph~(1b), which gives the crucial $m^2_N$-depedence in eq.~(29),
is absent when the $Z^0$~boson is replaced by a photon.\\

It is also important to notice that similar strong mass depedence of the
transition amplitude (i.e.~eq.~(29)) has previously been observed
in the quark sector in $Z^0$~decays~[13], e.g.~in $Z^0 \to b\bar{b}$,
$Z^0\to b\bar{s}$,
where the top quark plays the role of the heavy neutrinos.
This strong mass depedence will be a common feature for theories based on
the spontaneous symmetry breaking mechanism. In ref.~[8],
the possibility of lepton-flavor-changing $Z^0$ decays was studied
in a superstring  inspired $SM$. The enhancement effect due to
intermediate heavy neutrinos should be also present in that model, if terms
proportional to $\zeta^2\ m^2_N/M^2_W$ are not neglected and the whole
pertubatively allowed parameter space is considered.\\

In conclusion, we have explicitly
demonstrated that $SU(2)\otimes U(1)$ models with more than one
right-handed neutrino can theoretically account for {\em sizeable}
lepton-flavor-nonconservation effects at the $Z^0$~peak. Another
attractive theoretical feature of these models
is that {\em such rare leptonic $Z^0$ decays
are generally not suppressed by the usual "see-saw" mechanism}. Obviously,
an improvement of the current experimental limits on possible
non-universality effects in the tau lifetime~[14] and on
$Br(Z^0\to \tau e$ or $\tau \mu$) provides a possibility of
imposing upper bounds on the masses of the heavy Majorana neutrinos.\\
\bigskip

\noindent
{\bf Acknowledgements.} One of us (A.P.) wishes to thank T.~Yanagida,
G.~Zoupanos, Z.~Berezhiani, B.~Kniehl and C.~Greub
for useful conversations during the DESY workshop (1992),
where a relevant talk on this issue was
given. We also thank M.~M.~Tung for a critical reading of the manuscript.

\newpage

\centerline{\bf\Large Figure and Table Captions }
\vspace{1cm}
\newcounter{fig}
\begin{list}{\bf\rm Fig. \arabic{fig}: }{\usecounter{fig}
\labelwidth1.6cm \leftmargin2.5cm \labelsep0.4cm \itemsep0ex plus0.2ex }

\item Feynman graphs responsible for the effective $Z^0-l_1-l_2$ coupling
($l_1\neq l_2$) in the Feynman--'t Hooft gauge.

\item Theoretical bounds on the mass of the heavy Majorana neutrino $m_N$
and the $CKM$ mixing combination $\zeta=B_{l_1N}B^\ast_{l_2N}$ in a two
generation model. These bounds result from the following requirements:
(i)~$Br(Z^0 \to \bar{l}_1l_2+\l_1\bar{l}_2) \leq 10^{-5}$ (solid line),
(ii)~$Br(Z^0 \to \bar{l}_1l_2+\l_1\bar{l}_2) \leq 10^{-6}$
(dashed line),
(iii)~the validity of pertubative unitarity (i.e.~$\Gamma_N / m_N \leq 1/2$)
(dash-dotted line). The area lying to the right of the curves is not
allowed due to the conditions mentioned above.
\end{list}
\newcounter{tab}
\begin{list}{\bf\rm Tab. \arabic{tab}: }{\usecounter{tab}
\labelwidth1.6cm \leftmargin2.5cm \labelsep0.4cm \itemsep0ex plus0.2ex }

\item  Numerical results of leptonic flavor-changing $Z^0$ decays in a two
generation model. We have further assumed that $m_{N_1} \simeq m_{N_2}$
and $\zeta = B_{l_1N_1}B_{l_2N_1}^\ast = 0.05$.

\end{list}

\newpage

\bigskip\bigskip\bigskip\bigskip\bigskip\bigskip

\centerline{\bf\Large Table 1}
\vspace{1.5cm}
\begin{tabular*}{13.44cm}{|rr|rr|}
\hline
 &&& \\
\hspace{1.cm} $m_{N}$~[TeV]& \hspace{1.cm} &  \hspace{2.cm}
$Br(Z^0 \to \bar{l}_1l_2+\l_1\bar{l}_2)$ &\hspace{2.cm} \\
&&& \\
\hline\hline
&&& \\
0.5&& 3.4~$10^{-8}$& \\
&&& \\
1. && 3.4~$10^{-7}$&  \\
&&& \\
2. && 3.6~$10^{-6}$& \\
&&& \\
3. && 1.6~$10^{-5}$& \\
&&& \\
4. && 4.7~$10^{-5}$& \\
&&& \\
5. && 1.1~$10^{-4}$& \\
&&& \\
6. && 2.3~$10^{-4}$& \\
&&& \\
\hline
\end{tabular*}

\end{document}